\documentclass{ws-procs975x65}

\begin{document}

\title{\Large Epicyclic Frequencies and Resonant Phenomena Near Black Holes: The Current
Status}

\author{ALIKRAM  N. ALIEV\footnote{aliev@gursey.gov.tr}}

\address{Feza G\"ursey Institute, P.K. 6  \c Cengelk\" oy, 34684 Istanbul, Turkey}

\begin{abstract}
We briefly discuss the theory of epicyclic motion near rotating
black holes and its relation to the origin of high frequency
Quasi-Periodic Oscillations (QPOs) seen in many cases of accreting
black hole binaries. We also point out some new  frequency ratios
predicted by the theory.
\end{abstract}
\bodymatter

In the framework of general relativity the successive theory of
epicyclic motion around rotating black holes was first developed
in  Refs.\cite{ag1,ag2}(see also a review paper \cite{agr}). As is
known, the rotating black holes are uniquely described by the
famous Kerr solution
\begin{equation}
ds^2  =  -{{\Delta}\over {\Sigma}} \left(dt - a \sin^2\theta\,
 d\phi\,\right)^2 + \Sigma \left(\frac{dr^2}{\Delta} + d\theta^{2}\right) + \frac{\sin^2\theta}{\Sigma}
\left[a dt - (r^2+a^2) \,d\phi \right]^2 ,\label{kmetric}
\end{equation}
where the metric functions
$$ \Delta= r^2 +
a^2 -2 M r \,\,,~~~~~~\Sigma=r^2+a^2 \cos^2\theta \,.
$$
The equation of motion for a test particle of mass $ \mu $ and
charge $ e $  has the form
\begin{equation}
\frac{d^2 x^{\mu}}{ds^2}+ \Gamma^{\mu}_{\alpha \beta} \frac{d
x^{\alpha}}{ds}  \frac{d x^{\beta}}{ds}= \frac{e}{\mu}\,
F^{\mu}_{~~\nu}\, \frac{d x^{\nu}}{ds}\,\,. \label{eqmot}
\end{equation}
From the symmetry properties of the metric (\ref{kmetric}) it
follows that a cyclic motion of the test particle  is possible in
the equatorial plane $\,\theta=\pi/2\,$. For the cyclic motion we
have $ z^{\mu}(s)=u^{0} s \,\, \{1,0,0,\omega_0 \} $, where  $
\omega_0 $ is  the angular velocity of the motion.

To describe the epicyclic motion of the test particle we employ  a
perturbative expansion  about the cyclic orbits, i.e. we assume
that $ x^{\mu}(s)= z^{\mu}(s)+\xi^{\mu}(s) $. Substituting this
into equation (\ref{eqmot}) and adding also the external force
$f^{\mu}=f^{\mu}_{~~\nu} u^{\nu} (u^{0})^{-1} $, we arrive at the
equation
\begin{equation}
\frac{d^2 \xi^{\mu}}{dt^2} + \gamma^{\mu}_{\alpha}\,\frac{d
\xi^{\alpha}}{dt} + \xi^a \partial_a U^{\mu} =\frac{e}{\mu u^0}
\,f^{\mu} + {\cal N^{\mu}}\left(\xi, \frac{d
\xi}{dt}\right)\,,~~~~a=1,2\equiv r,\theta \label{perteq}
\end{equation}
where  $ {\cal N^{\mu}} $ stands for the non-linear terms, the
quantities   $ \gamma^{\mu}_{\alpha} $ and $
\partial_a U^{\mu} $ must be taken on a cyclic orbit  $r=r_0,\,\, {\theta=\pi/2}
$ and
\begin{eqnarray}
\gamma^{\mu}_{\alpha}&=& 2 \Gamma^{\mu}_{\alpha \beta}
u^{\beta}(u^{0})^{-1} - \frac{e}{\mu u^0} \,F^{\mu}_{~\alpha}\,,~~
U^{\mu}= \frac{1}{2} \left[ \gamma^{\mu}_{\alpha}u^{\alpha}
(u^{0})^{-1} - \frac{e}{\mu u^0} \,F^{\mu}_{~\alpha}
\,u^{\alpha}(u^{0})^{-1} \right]\, \label{pertquant}
\end{eqnarray}
Performing an appropriate integration  in (\ref{perteq}) and
restricting ourselves to the first order  terms in $ \xi $ we
obtain the set of equations
\begin{eqnarray}
\frac{d \xi^{A}}{dt} + \gamma ^A_{1}\, \xi ^r &=&  \frac{e}{\mu
u^0}\int f^A dt\,\,,~~~~~~~~~~A=0,3\equiv
t,\,\, \phi  \\[3mm]
\frac{d^2 \xi^{\,r}}{dt^2} +\omega_{r}^2 \xi^{\,r}&=& \frac{e}{\mu
u^0}\left( f^{\,r} - \gamma^{1}_A \int f^A dt \right) \,\,, \\[3mm]
\frac{d^2 \xi^{\,\theta}}{dt^2} +\omega_{\theta}^2 \xi^{\,\theta}
&=& \frac{e}{\mu u^0}\, f^{\,\theta}\,\,.
\end{eqnarray}
We see  that in the absence of the external force the epicyclic
motion is governed by the two decoupled oscillatory-type
equations. In other words, it amounts to free radial and vertical
oscillations with the associated frequencies
\begin{eqnarray}
\omega_{r}&=& \left(\frac {\partial U^{r}}{\partial r}-
\gamma_{A}^1\, \gamma_{1}^A
\right)^{1/2}\,\,,~~~~~~\omega_{\theta}= \left(\frac {\partial
U^{\theta}}{\partial \theta}\right)^{1/2} \,.\label{gfreqs}
\end{eqnarray}
The explicit expressions for these frequencies were first found in
Ref. \cite{ag1} in the most general case of  black holes with an
electric charge or with an external uniform magnetic field.
Further, the epicyclic frequencies were used to explore the maser
effect near black holes \cite{ag2} as well as the  effects of
periastron precession and the Lense-Thirring  drag in the
Schwarzschild and Kerr fields threaded by a cosmic string
\cite{ag3}. For a "pure" Kerr field the frequencies are
\cite{ag1,ag2}
\begin{eqnarray}
\omega_{r}^2 &= &\omega_{0}^2\, \left( 1-\frac{6 M}{r} -\frac{3
a^2}{r^2} \pm \, 8 a \omega_{K}
\right)\,\,,~~~~~\omega_{\theta}^2= \omega_{0}^2\, \left(1
+\frac{3 a^2}{r^2} \mp \, 4 a \omega_{K} \right)\,\,,
\label{radaxial}
\end{eqnarray}
where $\omega_0= \pm \,\omega_K/(1 \pm a \omega_K) $, $\omega_K
=\sqrt{M/r^3} $ is the Kepler frequency and  the two signs refer
to direct and retrograte motions. Later, in 1987  our expression
for the frequency of radial oscillations was re-derived in
Ref.\cite{kato1}, while the formula for the frequency of vertical
oscillations re-appeared \cite{kato2} only in 1990.

In Ref. \cite{ag1} we have also pointed out that when taking into
account the non-linear terms on the right-hand-side of Eq.
(\ref{perteq}) a coupling between the oscillations  occurs. If the
frequencies $ \omega_r  $ and  $ \omega_{\theta} $ are in the
rational relation $ k_r\, \omega_r \,= k_{\theta}\,
\omega_{\theta}  $, where $ k_r $ and  $ k_{\theta} $ are
integers, the resonances must take place. We have  plotted the
positions of low-order  $(k= \mid k_r \mid+ \mid k_{\theta} \mid)$
resonances: $ k=3,\, 4,\, 5, ~~~~or~~~( 2:1,\, 3:1,\, 3:2 )$. In
recent developments, this theory has been extensively used to
explain the frequency ratios and the origin of high frequency
Quasi-Periodic Oscillations (QPOs) observed in many cases of
accreting black hole or neutron star binaries (see papers
\cite{mvs, taks} and  references therein). In some cases, even our
plots  for the positions of the non-linear resonances were exactly
reproduced (see for instance, Ref.\cite{marek}).

Our model also admits the appearance of the forced resonances
under a perturbing force. Solving  Eqs.(5)-(7) for  Fourier
transforms we obtain
\begin{eqnarray}
\xi^{A}(\omega,m) &=&  -\frac{i}{\omega_m}\,\gamma^{A}_{1}
\,\xi^{\,r}(\omega,m) - \frac{e}{\mu u^0}\,\frac{f^{A}(\omega,
 m)}{\omega_{m}^2} \,\,, \\[2mm]
\xi^{\,r}(\omega, m) &=& \frac{e}{\mu
u^0}\,\frac{f^{\,r}(\omega,m)-\frac{i}{\omega_m}\, \gamma^{1}_{A}
f^{\,A}(\omega,m)}{\omega_{r}^2- \omega_{m}^2} \,\,,\\[2mm]
\xi^{\,\theta}(\omega, m) &=& \frac{e}{\mu
u^0}\,\frac{f^{\,\theta}(\omega,m)}{\omega_{\theta}^2-
\omega_{m}^2} \,\,,~~~~~~\omega_{m}= \omega- m \,\omega_{0}\,\,.
\label{fsol}
\end{eqnarray}
We see that at the frequencies $ \omega= m \omega_{0}\,,~~
\omega=\Omega_{r}^{\pm}= m \omega_{0} \pm \omega_{r}\,\,,~~
\omega=\Omega_{\theta}^{\pm}= m \omega_{0} \pm \omega_{\theta}\,\,
$  the system exhibits the resonant behaviour \cite{ag2}. The
similar type of the resonant frequencies was discussed by Kato
\cite{kato3} within a model  for kHz QPOs resonantly induced in
warped disks of X-ray binaries. In the latest developments, we
have performed a complete numerical analysis of the frequency
ratios pertaining to the orbits where the frequency of the radial
oscillations approaches its maximum value. Some results for direct
orbits and for masses $\thicksim 10 M_{\odot} $ are given in the
Table.

\hspace{-0.2 in}\begin{tabular}{cccccccc} \hline
$ a/M$&$r_{max}/M$&$\nu_{0} $(Hz)$ $&$\nu_{r} $(Hz)$ $&$\nu_{\theta} $(Hz)$ $&$\nu_{\theta}/\nu_{r}$&$\nu_{0}/\nu_{r}$&$\nu_{0}/\nu_{\theta}$\\
\hline
0.00 & 8.00 & 141.42 & 70.71 & 141.42 & 2.00 & 2.00 & 1.00\\
0.10 & 7.58 & 152.60 & 75.73 & 151.17 & 2.00 & 2.02 & 1.01\\
0.20 & 7.15 & 165.68 & 81.52 & 162.37 & 1.99 & 2.03 & 1.02\\
0.30 & 6.70 & 181.26 & 88.27 & 175.45 & 1.99 & 2.05 & 1.03\\
0.40 & 6.24 & 199.97 & 96.29 & 190.74 & 1.98 & 2.08 & 1.05\\
0.50 & 5.76 & 223.29 & 105.99 & 209.23 & 1.97 & 2.11 & 1.07\\
0.60 & 5.26 & 252.90 & 118.03 & 231.78 & 1.96 & 2.14 & 1.09\\
0.70 & 4.71 & 292.71 & 133.53 & 260.59 & 1.95 & 2.19 & 1.12\\
0.80 & 4.11 & 349.83 & 154.57 & 298.90 & 1.93 & 2.26 & 1.17\\
0.90 & 3.42 & 442.93 & 185.95 & 353.95 & 1.90 & 2.38 & 1.25\\
0.99 & 2.45 & 662.11 & 243.45 & 447.52 & 1.84 & 2.72 & 1.48\\
\hline
\end{tabular}

\vspace{3mm} \noindent {\em Acknowledgements:} The author thanks
the Scientific and Technological Research Council of Turkey
(T{\"U}B\.{I}TAK) for partial support under the Project 105T437.

\vfill
\end{document}